\documentclass[aps,prl,twocolumn,showpacs,floatfix]{revtex4}
\usepackage{epsfig}
\usepackage{psfrag}
\begin{document}
\title{Structure formation in binary colloids}

\author{ I.\ Varga${}^1$, F.\ Kun${}^{1}\footnote{Electronic 
address:feri@dtp.atomki.hu}$, K.\ F.\ P\'al${}^2$}
\affiliation{
${}^1$Department of Theoretical Physics,
University of Debrecen, P.\ O.\ Box:5, H-4010 Debrecen, Hungary \\
${}^2$Institute of Nuclear Research (ATOMKI), P.\ O.\ Box 51, H-4001
Debrecen, Hungary} 
\date{\today}
   
\begin{abstract}
A theoretical study of the structure formation observed very recently
[Phys.\ Rev.\ Lett.\ {\bf 90}, 128303 (2003)]  in binary colloids is
presented. In our model solely the dipole-dipole interaction of the
particles is considered, electrohidrodynamic effects are excluded.  
Based on molecular dynamics simulations and analytic
calculations we show that the total concentration of the particles,
the relative concentration and the relative dipole moment of the
components determine the structure of the colloid. At low
concentrations the kinetic aggregation of particles results in fractal
structures 
which show a crossover behavior when increasing the
concentration. At high concentration various lattice structures are
obtained in a good agreement with experiments. 
\end{abstract}
 

\maketitle
 
Recently, the structure formation in a monolayer of colloidal
particles under an 
external electric or magnetic field attracted considerable scientific
interest. Investigations have been performed with several different
types of particles subjected to a constant or time dependent external
fields. While in the absence of an external field chains, rings and
fractal clusters have been observed with a
temperature dependent fractal dimension
\cite{skjeltorp,tavares1,tavares2,teixeira,ghazali,huang1,stabil3,wen1,kun,blair,stramba},
in the presence of an external field crystal 
structures emerged \cite{wen_cryst,crystal1,crystal2}. It was
demonstrated that by 
appropriate adjusting of the external field all the planar lattice
structures can be produced in a magnetic monolayer \cite{wen_cryst}, 
which is of high practical importance and also addresses essential
problems of statistical physics like the study of phase transitions
between different morphologies and the phenomenon of melting in two
dimensions \cite{wen_cryst,crystal1,crystal2}. 

Binary colloids, {\it i.e.} suspensions composed of two sorts of
particles of the same size have been found very recently to produce
novel structures when subjected to an ac electric field
\cite{ristenpart}. Experiments 
performed with different frequencies and volume fractions of the
components revealed a
rich variety of self assembling structures in certain frequency
ranges: at frequencies where the effective particle interaction is
repulsive a triangular lattice of particles (Wigner crystal) was
observed \cite{ristenpart}.
At frequencies however where different types of particles attract each
other, aggregation and crystallization occurred: at low concentrations
strings and rings of alternating particles emerged while at higher
concentrations  various lattice structures were obtained
\cite{ristenpart}. 
It was found that the morphology 
also depends on the relative concentration of the particles, {\it
  i.e.} if one type of particles has 
a sizeable excess flower-like clusters are formed with one type of
particle in the middle surrounded by 5-6 particles of the other
type. In regions of $1:1$ or $1:2$ relative concentrations particles
crystallize into binary honeycomb or square-packed super-lattices. It
has been shown that electrohydrodynamic (EHD) flow is alone not enough
to understand the obtained structure formation but the dipolar
interaction of the particles arising due to their induced dipole
moment might have a governing role to produce the observed structures
\cite{ristenpart}.  
 
 In this paper we present a theoretical study of the novel structure
formation in binary colloids observed in Ref.\ \cite{ristenpart}.
Our goal is to understand the effect of dipolar forces on the
structure formation, 
furthermore, to identify the governing parameters of the system and
explore all possible aggregate and crystalline morphologies
determining the corresponding parameter regime of their occurrence.  

In the experiment the particles are settled down to the bottom plate
of a container and an ac electric field is imposed perpendicular to
the bottom. Under the influence of the external field the
particles attain a dipole moment parallel to the field, however, the
magnitude and the actual direction depends on the frequency of the
field and can be different for particles made of different materials.  
In our model an ensemble of $N$ particles of radius $R$ is considered
in a square box of side length $L$.
Point-like dipole moments are assumed to be placed in the middle of the
particles. For simplicity, the 
dipole moment of the particles is fixed during the time evolution of
the system to be perpendicular to the plane of motion pointing either
upward (particle type I) or downward (particle type II), which
represents the two different material properties of the
particles. This assumption implies that in the model only that
frequency range of the experiment is considered where different
particles attract each other.    
The particles are considered to be suspended in an
electromagnetically passive liquid  so that EHD effects are completely
excluded in the model. The liquid only exerts a friction force
(Stokes force) on the
particles which move under the action of dipole-dipole forces. When
particles touch each other during their motion we introduce a
repulsive force of the form of Hertz contact to prevent the overlap
\cite{wen1,kun}. After generating an initial configuration with random
particle positions in the simulation box, and fixing the dipole
moments according to the different material's properties, the time
evolution of the system is followed by solving numerically the
equation of motion of particles for the two translational degrees of
freedom with periodic boundary conditions 
(molecular dynamics simulation). In the model the motion of particles is
deterministic, {\it i.e.} no stochastic forces are taken into account
due to the high particle mass which hinders thermal motion
\cite{ristenpart}. For details of our simulation
techniques see Refs.\ \cite{wen1,kun}.  
\begin{figure}
\begin{center}
\psfrag{aa}{{\Large a)}}
\psfrag{bb}{{\Large b)}}
\psfrag{cc}{{\Large c)}}
\psfrag{dd}{{\Large d)}}
\epsfig{bbllx=0,bblly=0,bburx=400,bbury=390,file=ossz_aggreg.eps,
  width=7.0cm} 
 \caption{\small {\it (color)} The aggregation of dipoles with fixed
   $\phi_r = 1$, $\mu_r = 1$ at different
   concentrations: $a)$ $\phi = 0.075$, $b)$ $\phi = 0.125$, $c)$ $\phi
   = 0.5$, and $d)$, $\phi = 0.70$. For each simulation 1000 particles
   were used and $\phi$ was controlled by varying the side length
   $L$. 
} 
\label{fig:kin_aggreg} 
\end{center} 
\end{figure}  
A model system of $N$ particles consists of $N_1$ and $N_2$ particles
of dipole moment $\mu_1$ pointing upward $(+)$, and dipole moment
$\mu_2$ pointing downward $(-)$, respectively. The concentration of
particles $\phi$ is defined as the coverage $\phi = N R^2\pi/L^2$. The
partial concentration of the two components 
$\phi_1$, $\phi_2$ can be defined analogously whose ratio provides
their relative concentration $\phi_r = \phi_2/\phi_1$.
In the model the magnitudes $\mu_1$, $\mu_2$ of
dipole moments can be freely varied which captures up to some extent
the effect of frequency tuning of the experiment. For simplicity, we
fix $\mu_1$ and vary the ratio $\mu_r=\mu_2/\mu_1$. 
Two particles with parallel dipole moments exert a force onto each
other which is isotropic (central), it always falls
in the plane of motion parallel to the line connecting the two
particles. The force has a $1/r^3$ dependence on the separation $r$ of
the particles and it is repulsive for particles of the same type and
attractive for different ones. It is interesting to note that the
system is rather similar to an 
ensemble of charges where the interaction force decreases faster than
the Coulomb force. 

Simulations have been carried out varying the three parameters of the
model $\phi, \phi_r$ and $\mu_r$ in a broad range. 
Figure \ref{fig:kin_aggreg} shows simulation results obtained at different
concentrations $\phi$ fixing the relative concentration $\phi_r = 1$
and the relative dipole moment $\mu_r=1$. It can be seen that at
low concentrations the particles undergo a kinetic aggregation process
and form chains and rings of alternating types of particles due to the
attractive interaction of oppositely directed dipoles (Fig.\
\ref{fig:kin_aggreg}$a$ and $b$). It is 
interesting to note that the isotropic interparticle forces gives rise
to anisotropic clusters since already a cluster of two particles
behaves as an extended dipole oriented in the plane of motion which
then results in a favorable direction for the later
aggregation steps. At low concentrations the chain is the dominating
morphology, however, it can be shown analytically that for an even
number of alternating particles the ring structure is energetically
favorable \cite{preprint1}. (A ring can be observed in the upper part
of Fig.\ \ref{fig:kin_aggreg}$b$.) The length of alternating chains
is limited, 
when the chain length becomes comparable to the average distance of
chains, aggregation can occur not only at chain ends but also at
internal particles resulting in branching and more compact
structures. In order to 
characterize the structure of growing clusters we calculated the
radius of gyration $R^2_g(n) = 1/n(n-1)\sum_{i\neq
  j}(\vec{r}_i-\vec{r}_j)^2$ for each cluster during the time 
evolution of the system and averaged over clusters of the same size
$n$. 
\begin{figure}
\begin{center}
\epsfig{bbllx=145,bblly=370,bburx=475,bbury=670,
file=fractal.ps, width=8.5cm} 
 \caption{\small  The number of particles $n$ of clusters
   as a function of the radius of gyration $R_g$. The dashed and
   continuous straight lines have slope 1 and 1.45, respectively. The
   crossover value $n_c$ was obtained as the crossing point of the
   straight lines fitted to the two regimes of different slopes. The
   inset shows $n_c(\phi)$ where a straight line of slope $1.18\pm
   0.05$ was fitted to the data.  
  } 
\label{fig:fractal} 
\end{center} 
\end{figure}
In Fig.\ \ref{fig:fractal} the cluster size $n$ is plotted as a
function of the averaged gyration radius $R_g$
on a double logarithmic plot for several different concentrations. It
can be observed in the figure that for each concentration
$n(R_g)$ is composed of two parts of power law functional
form $n \sim R_g^{\alpha}$: for small clusters
$n(R_g)$ can be well fitted by a power law of exponent close to $1$
indicating chainlike structures. 
At a certain chain length $n_c(\phi)$, however, a
crossover occurs into a more compact structure characterized by a
higher exponent of $R_g$, $\alpha = 1.45 \pm 0.05$ was determined from
simulations for $\mu_r = 1$ and $\phi_r = 1$. It follows from our
argument that the 
crossover chain length $n_c$ has a power law dependence on the
concentration, {\it i.e.}  $n_c \sim \phi^{-\beta}$ with an exponent
$\beta = 1.0$. The numerical result is presented on the inset of Fig.\
\ref{fig:fractal} where $\beta = 1.18 \pm 0.05$ was obtained in a
reasonable agreement with the analytic prediction. 
Simulations also showed that the exponent $\alpha$ characterizing the
structure of growing aggregates does not depend on $\phi$ until $\phi
< \phi^{*} \approx 0.25$. It is important to emphasize however that
$\phi_r$ and $\mu_r$ have a substantial effect on the aggregation
process which will be explored in its entire complexity elsewhere
\cite{preprint1}. 
 
Increasing the concentration above $\phi^{*}$ the morphology of
clusters changes 
drastically; crystallites of square-packed structure are formed which
then aggregate into large clusters. 
 Fig.\ \ref{fig:kin_aggreg}$c$ shows simulation results obtained at $\phi =
 0.5$ where a network of crystallites can be observed. At higher
 concentrations the entire system self assembles into a square-packed
 lattice of alternating particles as in Fig.\ \ref{fig:kin_aggreg}$d$.

In order to understand the formation of crystal structures in binary
colloids and explore possible lattice morphologies beyond the
square-packed one, we calculated the energy of basic particle
configurations from which lattices can be built up. The Basic
Configurations (BC) are composed of a particle of type I (with dipole
moment $\mu_1$) surrounded by
2 ($BCI$), 3 ($BCII$), 4 ($BCIII$), 5 ($BCIV$), or 6 ($BCV$) particles
of type II (with dipole moment $\mu_2$) placed at the energetically
most favorable locations. In 
order to simply cover also cases where the different types of particles are
interchanged, the energy of BCs was calculated as a function of
$\mu_r=\mu_2/\mu_1$ fixing the value of $\mu_1 = 1$ 
\begin{eqnarray}
E_N(\mu_r) = \frac{\mu_1^2}{d^3}\left[\frac{\mu_r^2}{8}\sum_{k=1}^{N-2}
\frac{N-k-1}{\sin^3(k\beta)} -\mu_r(N-1) \right],
\label{eq:ener_bc}
\end{eqnarray}
where $\beta = \pi/(N-1)$ and the particle number takes the values
$N=3,4,5,6,7$. The energy divided by the particle number 
$E_N(\mu_r)/N$ is presented 
in Fig.\ \ref{fig:basic_conf}  as a function of $\mu_r$ for the
possible values of $N$. 
It can be seen in the figure that all the BCs are stable (have
a negative energy) only in certain $\mu_r$ ranges between $0$ and
a $\mu_r^{max}(N)$ since with increasing $\mu_r$ the attraction exerted
by the central particle is not enough to compensate the mutual
repulsion of the surrounding ones. 
Such configurations where the central particle is surrounded by
particles of smaller dipole moment, {\it i.e.} $\mu_r < 1$ in Fig.\
\ref{eq:ener_bc}, are always stable except for $BCV$ because
$\mu_r^{max}(7) \approx 0.8$ falls below 1. The upper bounds can be
determined exactly from Eq.\ (\ref{eq:ener_bc}) as $\mu_r^{max}(N) \approx
16, 5.2, 2.4, 1.3, 0.8$ was obtained for $N=3,4,5,6,7$, respectively.
\begin{figure}
\begin{center}
\psfrag{aa}{BCI}
\psfrag{bb}{BCII}
\psfrag{cc}{BCIII}
\psfrag{dd}{BCIV}
\psfrag{ee}{BCV}
\epsfig{bbllx=0,bblly=-5,bburx=570,bbury=470,
file=ener_bind_uj.eps, width=8.5cm} 
 \caption{\small  
$E_N(\mu_r)/N$ from Eq.\ (\ref{eq:ener_bc}) 
for $N=3,4,5,6,7$. For clarity, the corresponding particle configurations
are also presented.
  }
\label{fig:basic_conf} 
\end{center} 
\end{figure}
Repeating the above basic structures with alternating particles
various structures can be built up, {\it i.e.}  $BCI$ results in
chains or rings, $BCII$ gives rise to a 
honeycomb lattice in which both types of particles have three
neighbors of the other type, and $BCIII$ leads to 
the square lattice. $BCIV$ 
results only in flower-like structures, no lattice can be constructed
since the plane cannot be covered by regular pentagons.  Moreover,
$BCV$ forms the basis of a special type of honeycomb lattice, in which
particles of the larger dipole moment have $6$ neighbors of the
smaller one but particles of the smaller moment have 3 neighbors of
both types.
\begin{figure}
\begin{center}
\psfrag{aa}{{\Large a)}}
\psfrag{bb}{{\Large b)}}
\psfrag{cc}{{\Large c)}}
\psfrag{dd}{{\Large d)}}
\psfrag{ee}{{\Large e)}}
\psfrag{ff}{{\Large f)}}
\epsfig{bbllx=0,bblly=0,bburx=660,bbury=410,file=uj_ossz_kristaly.eps,
  width=8.5cm} 
 \caption{\small {\it (color)} Aggregate and crystalline morphologies
   obtained by 
   simulations. In each system 1000 particles were used and
   concentration was controlled by changing the size of the
   simulation box. For visualization smaller pieces of the simulation
   box are cut out and magnified. 
}
\label{fig:lattices}
\end{center}
\end{figure}
The overall morphology attained by the entire particle system at a
given 
value of $\mu_r$ is not necessarily based on the energetically most 
favorable basic configuration. In spite of the isotropic
particle-particle interaction the cluster-cluster interaction is
highly anisotropic which can result in trapping the particles into
local energy minima in configuration space, especially when thermal
motion is hindered by the relatively large particle mass. Hence, the
relative dipole moment $\mu_r$, the
total concentration $\phi$ and the relative concentration $\phi_r$ of
components together determine the final structure.
Based on the energy of basic configurations Eq.\ (\ref{eq:ener_bc})
and Fig.\ \ref{fig:basic_conf}, it is possible to determine analytically
regions of $\phi, \phi_r$, and $\mu_r$ which are needed to obtain a
certain lattice structure or aggregate morphology. 
The numerical results are summarized in Table \ref{tab:table1} and
Fig.\ \ref{fig:lattices} provides an overview of possible
structures obtained by simulations varying $\phi, \phi_r, \mu_r$. 

If one of the components has a much larger concentration 
than the other one (for instance, $\phi_1 >> \phi_2$), due to the
repulsive interaction of the identical particles a triangular lattice
is formed with some binary islands. The limiting case when only one of
the components is present is shown in Fig.\ \ref{fig:lattices}$a$
where a regular triangular lattice is observed. The
simplest basic configuration BCI gives rise to chain and ring
conformations which occurs at $\mu_r = 1, \phi_r = 1$, in the
concentration range $\phi < \phi^{*} \approx 0.25$ with a crossover
into more compact morphologies as discussed above. 
For an even number of particles the ring is energetically more
favorable than the chain configuration. In spite of this, chains cannot
close to form rings because the bending would require energy. Rings
appear when two nearby chains merge which occur most frequently at
concentrations $\phi \approx 0.125$, see Fig.\
\ref{fig:lattices}$b$). Similar chain and string formation of
alternating particles has been reported in Ref.\ \cite{ristenpart}. 

At high concentrations lattice structures can
be obtained, however, concentrations too close to the highest coverage
($\phi \approx 0.9$) are disadvantagous because they prevent particle
motion and result in freezing into local energy minima. Hence, the
best square lattice structures can be achieved at $\phi \approx
0.65-0.75$. This type of lattice contains the same amount of both type
of particles so that $\phi_r = 1$ and $\mu_r = 1$ follows, see Fig.\
\ref{fig:lattices}$c$. The parameter regime providing square lattice
agrees well with the experimental observation of Ref.\
\cite{ristenpart}.   
\begin{table}
\begin{center}
\caption{ \label{tab:table1} The parameter regimes determining final
  structure of the system. For explanation, see the text. (In the
  first row missing values of $\phi_{r}$,  $\mu_r$ indicate that
  solely one of the components is present.)} 
\begin{tabular}{|c|c|c|c|}
\hline
\hline
$Structure$ & $\phi$ & $\phi_{r}$ & $\mu_r$  \\
\hline 
triangular latt.\ & $0 - 0.9$       & $-$              & $-$       \\
chain             & $0 - 0.125$     & $1$              & $1$       \\
ring              & $0.125$         & $1$              & $1$       \\
square lattice    & $0.65-0.75$     & $1$              & $1$       \\
honeycomb I       & $0.45-0.50$     & $1$              & $\approx 2.5$\\
honeycomb II      & $0.80-0.82$     & $2$              & $2.5-$    \\
super lattice     & $0.45-0.50$     & $\approx 2-2.5$  & $2.5-4.5$ \\ 
\hline
\end{tabular}
\end{center}
\end{table}
The first type of honeycomb lattice ({\it honeycomb I}) where each
particle has 3 neighbors of the other type has not been observed
experimentally. However, our analytic calculations showed that
energetically it can be favorable for the system in a certain
parameter range. An example is presented in Fig.\
\ref{fig:lattices}$d$ which was obtained at $\phi = 0.5$ and $\phi_r =
1$. For clarity, some plaquettes of the lattice are highlighted in the
figure. The basis of this structure is $BCII$ which is always stable
for $\mu_r < 1$, {\it i.e.} when a larger dipole is surrounded by 3
smaller ones. However, to make the inverse configuration more
favorable, $\mu_r$ has to be increased above the stability limit of
$BCIII$ (above $\mu_r^{max}(5) \approx 2.4$), otherwise, the system
ends up in a square packed structure.  $\mu_r \approx 2.5$ proved to
be an excellent choice numerically, see Fig.\ \ref{fig:lattices}$d$. 
If the magnitude of the two dipole moments are different ($\mu_r \neq
1$) concentration fluctuations easily lead to the formation of the
second type of honeycomb lattice  ({\it honeycomb II}), which has a
hexagonal-closed-packed (HCP) structure, where particles of the larger
dipole moment have 6 
neighbors of the other type and the ones with smaller dipole moment
have 3 of both types. The highest portion of the system was found to
crystallize into honeycomb II at $\phi \approx 0.80-0.82$ with the
ratio of $\phi_r = 2$ of the components and with high enough asymmetry of the
magnitude of dipole moments $\mu_r \approx 2.5$, which is needed to
prevent the system to crystallize locally into the square packed
structure. The corresponding simulation results can be seen in Fig.\
\ref{fig:lattices}$e$, which is in a nice agreement with the
experiments \cite{ristenpart}. If the relative dipole moment $\mu_r$
is even higher ( $\mu_r > 2.5$) but the concentration and the 
relative concentration do not favor the emergence of honeycomb II,
so-called super structures can be observed in the colloid. These
structures do not have ordered crystalline morphology, instead they are
characterized by long straight binary chains which connect disordered
or small crystalline island as in Fig.\ \ref{fig:lattices}$f$. 

In summary, we proposed a simple model of a binary monolayer of
dipolar particles to explore the effect of the induced dipole-dipole
interaction in the intriguing structure formation observed recently
\cite{ristenpart}. Varying the three parameters of the model, {\it
  i.e.} the total concentration of the particles $\phi$, the 
relative concentration $\phi_r$ and the relative dipole moment $\mu_r$
of the components a rich variety of structures were obtained in
satisfactory agreement with the experimental findings
\cite{ristenpart}. 
The simplicity of the model
demonstrates that the main qualitative features of the structure
formation observed are determined by the induced dipole-dipole 
interaction. Further theoretical studies are in progress and further
experiments are proposed.   

\begin{acknowledgments}
This work was supported by the projects OTKA T037212, T037991. F.\ Kun
was supported by the Research Contract FKFP 
0118/2001 and by the Gy\"orgy B\'ek\'esi Foundation of the Hungarian
Academy of Sciences. 
\end{acknowledgments}

\end{document}